\documentclass[aps,pre,twocolumn,groupedaddress,amsmath,amssymb,superscriptaddress]{revtex4-1}
\usepackage{graphicx}
\usepackage{subfigure}
\usepackage{hyperref}

\usepackage{epstopdf}

\usepackage{color}

\begin{document}
\date{\today}

\author{Jonathan Landy}
\email{landy@berkeley.edu}
\affiliation{Chemistry Department, University of California, Berkeley}

\date{\today}

\title{Shape-dependent bounds on cell growth rates}

\begin{abstract}
I consider how cell shape and environmental geometry affect the rate of nutrient capture and the consequent maximum growth rate of a cell, focusing on rod-like species like \textit{E.\ coli}.  Simple modeling immediately implies that it is the elongated profiles of such cells that allows for them to grow -- as observed -- at exponential rates in nutrient-rich media. Growth is strongly suppressed  when nutrient capture is diffusion-limited:  In three dimensions, the length is bounded by $\log L  \lesssim   t^{1/2}$, and in lower dimensions  growth is algebraic.  Similar bounds are easily obtained for other cell geometries, groups of cells, \textit{etc}.  Fits of experimental growth curves to such bounds can be used to estimate various quantities of interest, including generalized metabolic rates.
\end{abstract}

\maketitle

\section{Introduction}
\begin{figure}[b]\begin{center}\scalebox{.45}
{\includegraphics[angle=0]{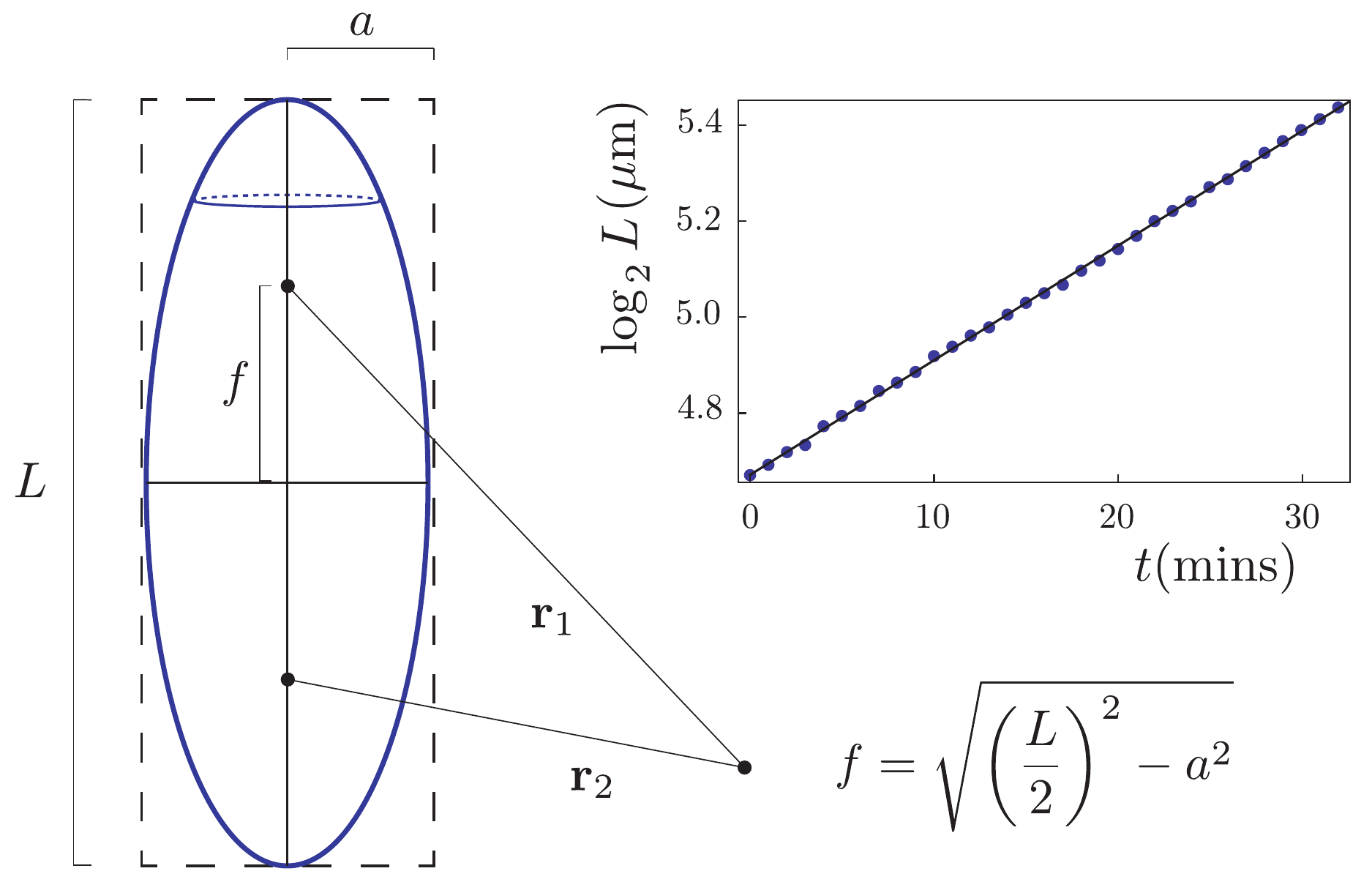}}
\caption{\label{fig:ellipse}  (color online) Idealized rod-like cell of length $L$ and radius $a$.  An ellipsoidal cell model is easy to treat analytically, and  returns a growth curve bound that is qualitatively similar to that of a cylindrical cell. For each point in space,  vector $\textbf{r}_i$  extends from focus $i$ to the point in question.  The surface $\vert\textbf{r}_1\vert + \vert\textbf{r}_2\vert \equiv L$ defines the cell boundary.  Inset: Part of an experimental filamentous \textit{E.\ coli} growth curve together with exponential fit (typical cell radii are $\approx 0.5 \mu$m) -- data from \cite{ami-13}. }
\end{center}
\end{figure}

It is fairly common knowledge that, given favorable  conditions, the population of a bacterial culture will grow at an exponential rate \cite{Mon-49, hag-10,scott2011bacterial}.  Less well known is the fact that the members of a healthy  culture often also individually grow exponentially fast:  Some species of bacteria are roughly cylindrical in shape, and when one of these is tracked over time, one finds that its radius $a$ is roughly fixed, but its length $L \equiv L(t)$ is exponential in form  --  cf.\ Fig.\ \ref{fig:ellipse} -- \cite{koch2001bacterial}.  This is true both for cells that divide once they have grown for some time \cite{Wan-10}, and also for filamentous cells that do not divide \cite{koch2001bacterial, ami-13}.  In this latter case, single-cell tracking often shows convincing, steady exponential growth over multiple decades in length. Exponential growth has also been  observed in budding yeast \cite{di2007effects}, which suggests that this  trait  can be exhibited by many, if not all, rod-like cells.

Here, I attempt to rationalize -- and also to determine the conditions that permit -- the observed  exponential growth rates of rod-like  cells.  To do so, I consider the effect of cell shape and environmental geometry on the rate of nutrient capture by a cell.  Focusing on isolated individuals, I point out that energy conservation immediately implies that it is the elongated profiles of rod-like cells that allows for their exponential growth in nutrient-rich media.  However, when nutrient capture is diffusion-limited, different growth bounds result: In this case, in three dimensions, nutrient conservation returns the upper bound, $\log L \lesssim t^{1/2}$, while in one and two dimensions cell length is even more severely restricted, being bounded by nearly-linear, power-law forms in time.  As I detail below, these latter results enable experimental extraction of quantitative measures relating to internal cell functionality.  For example, fits of diffusion-limited  growth data to the forms I derive here would return estimates of the nutrient-specific cellular metabolic rate, which is a quantity that would likely be difficult to probe using other experimental designs.

\section{Energy conservation \& abundant nutrient}
\label{sec:scaling}
To be concrete, I will suppose for the moment that the growth-limiting nutrient provides the cell's sole source of energy; all other materials needed for cell functionality I take to be available to the cell in  abundance.  In this case, to bound the rate of cellular growth, one need only specify three fundamental parameters:  1)     $\mathcal{P}_m$, the metabolic rate, which is the amount of energy utilized per unit time for upkeep of the cellular mass already in place, 2) $E_0$, the energy required to construct a unit volume of new cellular material, and 3) $\mathcal{P}_{in}$, the influx power, or rate at which energy is captured by the cell.  I suppose $\mathcal{P}_m$ to be an extensive quantity, proportional to the cellular volume, and $E_0$  a species specific constant.  Different experimental protocols are captured through appropriate modeling of $\mathcal{P}_{in}$, the remaining, third parameter.

In this section,  I consider the case where $\mathcal{P}_{in}$ is proportional to the cell's surface area.   This scaling applies in the limit of high nutrient concentration:  In this case, the rate of capture will be reaction-limited -- perhaps dominated by the binding time of nutrients to receptors --  and  $\mathcal{P}_{in}$ will consequently be proportional to the cell's surface area, assuming a roughly constant  receptor surface density  \cite{zwanzig1991time}.  Alternatively, this scaling should also apply -- both at high and low nutrient concentrations -- in situations where a nutrient bath is flowed past the cell, as in a microfluidic experiment \cite{Wan-10}.  In this case, a doubling of cell area results in a doubling of its cross-section, and a consequent doubling of nutrient collision and collection. Now, by assumption, construction of new cellular material requires there to be energy available in excess of that needed to satisfy the metabolic constraint.  Under the current assumptions, persistent cell growth thus requires
\begin{eqnarray} \label{cond1}
\mathcal{P}_{in} - \mathcal{P}_{m} \equiv \alpha_{in}A - \alpha_m V \geq 0,
\end{eqnarray}
where $\alpha_{in}$ and $\alpha_{m}$ are some constant coefficients\footnote{For a physical cell, $\alpha_m$ might change over time.  However,  setting this to any value less than or equal to its smallest possible value -- corresponding to the minimal energy expenditure needed to maintain cell life -- will provide a valid upper bound.  For example, in order to derive the simple bounds (\ref{subexp}), (\ref{subexp2}), and (\ref{subexp3}), I set $\alpha_m$ to zero.  These bounds are always valid.} and $A$ and $V$ are the cell's surface area and volume, respectively. 

The simple condition (\ref{cond1}) is sufficient to understand why cells of different shape exhibit different growth behaviors.  The key point is that the scalings of the two terms in (\ref{cond1}) are different for different cell geometries: For example, for a growing spherical cell of radius $R \equiv R(t)$, $A \propto R^2$, while $V \propto R^3$.  The metabolic term thus always grows more quickly  with cell size  than does the influx term, indicating that persistent cell growth must eventually terminate for spherical cells.  
 In contrast, rod-like cells of fixed radius $a$ but increasing length $L$ have $A(t) \approx 2\pi a L(t)$ and $V(t) \approx \pi a^2 L(t)$, both proportional to the running cell size.  In this case, if (\ref{cond1}) holds at time $t =0$, it will also hold at all  later times.  Further, a running bound on the length of a rod-like cell can be obtained by supposing that all incoming power in excess of the metabolic needs of the cell is immediately converted into new cellular material\footnote{If the cell opts to not quickly expend all excess power influx on the creation of new material, the net, time-integrated energy that it harvests will decrease due to its resulting smaller size.}.  In this extreme case, the cell's length satisfies the equation
 \begin{eqnarray}\label{exp}
\partial_t L = \frac{\left \{ \alpha_{in} 2 \pi a  - \alpha_m \pi a^2  \right \}}{E_0} L \to 
L \leq L_0 e^{\frac{\gamma}{E_0}t},
\end{eqnarray}
where $\gamma \equiv \left \{\alpha_{in} 2 \pi a  - \alpha_m \pi a^2 \right \}$.   

Evidently  conservation of energy places dramatically different constraints on the growth behavior of spherical and rod-like cells:  Whereas this condition requires that spherical cells must always eventually terminate their growth, even in favorable conditions, rod-like cells that  grow in only one direction  have metabolic rates and influx powers that are both proportional to cell size.  This allows for persistent cell growth, provided  the time-independent parameter combination $\gamma \geq 0$:  Exponential growth of rod-like cells can occur even when environmental conditions are only modestly favorable!

\section{Diffusion-limited growth, three dimensions} \label{sec:QTreatment}
I turn now to a second important scenario: cell growth subject to diffusion-limited nutrient capture.  To model such situations,  I suppose the limiting nutrient is at some bulk concentration $c_0$ far from the cell, it diffuses about, and is captured upon contact by the target cell.  These dynamics result in a  scaling form for $\mathcal{P}_{in}$ differing from that considered in the previous section.  Assuming cell growth is sufficiently slow that the nutrient concentration can be considered approximately equilibrated at any given moment (cf.\ discussion section for justification), it's profile will satisfy the steady-state diffusion equation
\begin{eqnarray} \label{diffeqn}
\nabla^2 c = 0.
\end{eqnarray}
The nutrient concentration profile must further satisfy two boundary conditions: It must approach $c_0$ at infinity, and it must be exactly zero at the cell's surface.   Berg and Purcell have made a series of insightful comments about such diffusion-to-capture-type processes in biology \cite{Ber-77, Berg-93}.  In particular, they have pointed out that the idealized, second boundary condition above actually holds to a very good approximation for physical cells, provided a few percent or so of their total surface area is able to capture the nutrient in question.  

Assuming an ellipsoidal cell geometry, as in Fig.\ \ref{fig:ellipse}\footnote{ The Laplacian separates in spheroidal coordinates \cite{Arf-70}, which enables straightforward analytic treatment of (\ref{diffeqn}) under ellipsoidal boundary conditions.  The growth curves of ellipsoidal and cylindrical cells should scale identically. }, the solution to (\ref{diffeqn}) that satisfies the two boundary conditions noted above is (cf.\ appendix)
\begin{eqnarray}\label{exactc}
c = c_0  - \frac{Q}{2f} \log \left  ( \frac{1+\frac{2f}{\vert \textbf{r}_1\vert +\vert \textbf{r}_2 \vert}}{1-\frac{2f}{\vert \textbf{r}_1\vert +\vert \textbf{r}_2 \vert}}\right),
\end{eqnarray}
where the definitions of $f$ (the cell's focal length) and the $\textbf{r}_i$ are given in Fig.\ \ref{fig:ellipse}, and
\begin{eqnarray}\label{exactq}
Q =\frac{ 2 f c_0}{\log \left ( \frac{1 + \frac{2f}{L}}{1 -\frac{2f}{L}}\right)} = \frac{\mathcal{P}_{in}}{4 \pi D}.
\end{eqnarray}
To obtain the last relation here, between $\mathcal{P}_{in}$ and $Q$, I have employed Fick's law, which states that the local flux of nutrient is given by
\begin{eqnarray}\label{FL}
\textbf{J} =- D \nabla c,
\end{eqnarray}
where $D$ is the diffusion constant of the nutrient and $c$ is its local concentration.   The total nutrient flux into the cell is given by the integral of the normal component of $\textbf{J}$ over the cell surface.  This can be evaluated using Gauss's law \cite{Arf-70}, giving  (\ref{exactq}).

I now use (\ref{exactq}) to bound the growth of rod-like cells, using the extreme equation
\begin{eqnarray}\label{correcteom}
\partial_t L  = \frac{\mathcal{P}_{in} - \mathcal{P}_{m}}{E_0} \equiv \frac{\gamma}{E_0} L.
\end{eqnarray}
First, let us consider the spherical limit, where $L \approx 2a$.  In this case, (\ref{exactq}) gives
\begin{eqnarray} \label{spherelim}
\mathcal{P}_{in}  =  4 \pi  D c_0 a \text{ \ \ \ \   (spherical limit).}
\end{eqnarray}
Assuming an initial cell shape that is nearly spherical, (\ref{spherelim}) sets the initial growth rate of rod-like cells.  Further, for spherical cells, (\ref{spherelim}) holds at all times.  Notice that (\ref{spherelim}) is explicitly not proportional to the area of the cell, but instead grows only linearly with $a$, the radius.  This can be understood through a consideration of Fick's law (\ref{FL}):  Although the total nutrient flux into the cell is given by an integral over the cell's surface, the integrand is proportional to $\nabla c$, which must scale as $\frac{c_0}{a}$, by dimensional analysis.  As a result, the total flux goes as $a^2 \times a^{-1} = a^1$, as given in (\ref{spherelim}) \cite{Ber-77, Berg-93}.  This has important consequences for spherical cells:  Whereas the threshold radius at which spherical cell growth must stop goes as the nutrient concentration to the first power in the flowing nutrient ensemble, it scales as the square root of nutrient concentration in the diffusion-to-capture ensemble.  This results in a cross-over effect, with diffusion-to-capture environments allowing for larger cell sizes at small nutrient concentrations, but smaller cell sizes in the opposite limit.

Now let us consider the extended limit for rod-like cells, where $L \gg a$.  In this case,  (\ref{exactq}) gives
\begin{eqnarray}\label{longlim}
\mathcal{P}_{in} \sim \frac{2 \pi  D c_0}{ \log\left(\frac{L}{a} \right)} L \text{\ \ \ \ \ (extended limit)}.
\end{eqnarray}
A form close to this could also have been  predicted using dimensional analysis:  Again, the total nutrient flux into the cell is given by the integral over its surface of $D \nabla c$.  The gradient of $c$ now changes from place to place on the cell's surface, but its typical  order of magnitude might be expected to scale as either $\frac{c_0}{a}$ or $\frac{c_0}{L}$.  The first of these returns the estimate $\mathcal{P}_{in} \approx 2 \pi a L \times \frac{D c_0}{a}  = 2 \pi D c_0 \times L$, which, when plugged into (\ref{correcteom}), results in predicted exponential growth.  This  estimate nearly agrees with the correct power influx (\ref{longlim}), but the two differ in that (\ref{longlim}) has a dimensionless, logarithmic term in its denominator.  This introduces a  monotonically decreasing $L$-dependence to the power influx per unit cellular length, $\frac{\mathcal{P}_{in}}{L}$.

Although the logarithmic factor  in  (\ref{longlim}) varies slowly with $L$, it turns out to significantly affect the growth behavior of the cell:   Neglecting the metabolic term, rearrangement of (\ref{correcteom}) and (\ref{longlim}) gives
 \begin{eqnarray}
 \frac{\log\left(\frac{L}{a} \right)}{L} \partial_t L  \equiv \frac{1}{2} \partial_t \left \{ \log\left(\frac{L}{a} \right) \right \}^2  \approx  \frac{2 \pi  D c_0}{ E_0},
\end{eqnarray}
which results in the bound
\begin{eqnarray}  \label{subexp}
L  \leq a \times \exp \left [ \left ( \frac{4 \pi  D c_0}{ E_0} t \right)^{1/2} \right ].
\end{eqnarray}
This indicates that persistent exponential growth is not possible when nutrient capture is diffusion-limited.  This conclusion also holds for growth near a flat substrate\footnote{If the system is bounded by a  substrate with normal $\hat{\textbf{n}} \equiv \hat{\textbf{n}}(\textbf{r})$, the boundary conditions become: 1) $c = 0$ a the cell surface, 2) $c \to c_0$ at infinity, and 3) $\nabla c \cdot \hat{\textbf{n}} = 0$ at the substrate boundary.  Assuming a flat substrate, the solution to (\ref{diffeqn})  is simply obtained by the method of images \cite{Arf-70}.  This shows that $\mathcal{P}_{in}$ for the single-cell/substrate geometry is the same as that in a cell plus mirror cell/no substrate geometry:  If the cell sits near the substrate's surface, the influx and growth bound in this latter geometry scale just as in (\ref{longlim}) and (\ref{subexp}), but with $\gamma \to \frac{\gamma}{2}$. }.

\section{Diffusion-limited growth, one and two dimensions}
\begin{figure}[t]\begin{center}\scalebox{.55}
{\includegraphics[angle=0]{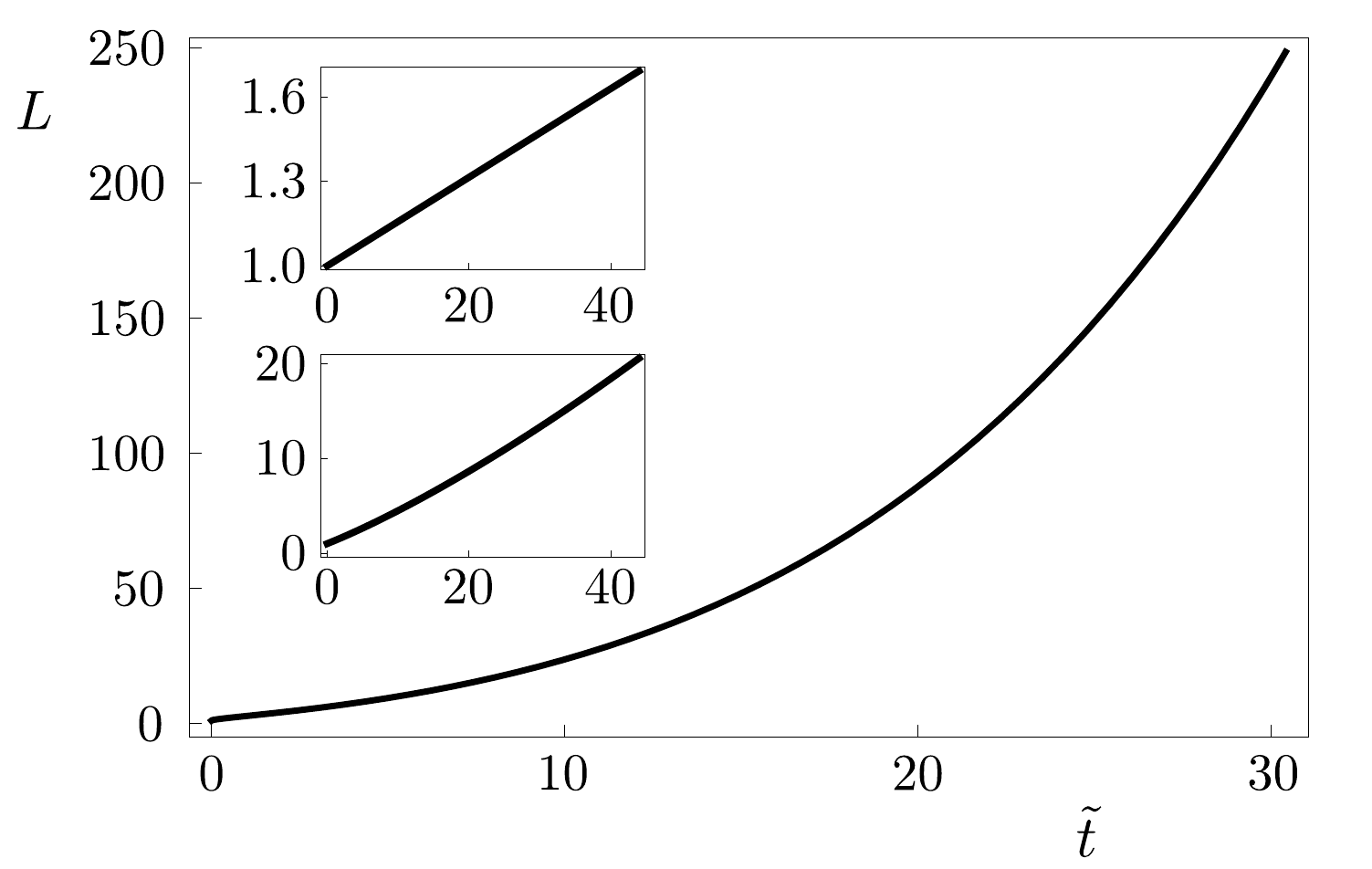}}
\caption{\label{fig:diffusive}  Diffusion-limited growth curves  for rod-like cells in 1-d [(\ref{subexp3}) with $R=100\ \mu$m, top inset], 2-d [(\ref{subexp2}) with $R = 100\ \mu$m, bottom inset], and 3-d [(\ref{subexp}), main plot] environments (length $L\  \mu$m \textit{vs.}\ time $\tilde{t} \equiv \frac{4 \pi D c_0 }{E_0} t$).  These are the long-length scale upper bounds that result when $\mathcal{P}_m$ is neglected.  Notice the differences in scale between the three plots: The nutrient capture rate is steadily decreased as the dimension is lowered.}
\end{center}
\end{figure}

 I now briefly consider  one and two dimensional diffusion-limited systems:   These examples illustrate the point that, in addition to a  strong dependence on cell shape and nutrient dynamics, cellular growth rates can also depend strongly on environmental geometry.  The two dimensional situation relates to an experimental setup where the cell and its surrounding solution are sandwiched between two slides, or cover-plates.  Far from the cell, at radius $R$, I take the limiting nutrient to have a fixed concentration $c_0$.  In this case, an analysis similar to the above returns the $L \gg a$ limiting form,
\begin{eqnarray} \label{twodinflux}
\mathcal{P}_{in} \sim \frac{2 \pi D c_0}{\log \left ( \frac{4 R}{L} \right)}.
\end{eqnarray}
Neglecting metabolic terms, this gives
\begin{eqnarray}
 \label{subexp2}
 L \left [1+ \log \left ( \frac{4R}{L}\right)  \right ] \leq  \frac{2 \pi D c_0}{E_0} t.
\end{eqnarray}
Growth is now bounded by a nearly-linear function in time, with a slope that increases slowly with $L$ -- cf.\ Fig.\ \ref{fig:diffusive}. 

Similarly, when the cell is placed in the center of a one-dimensional  channel of length $2 R$, we obtain
\begin{eqnarray}\label{one-dres}
\mathcal{P}_{in} = \frac{4 D c_0}{2 R-L},
\end{eqnarray}
where $c_0$ is the concentration at $x = \pm R$.
Neglecting metabolic terms here gives
\begin{eqnarray} \label{subexp3}
L \left (2 R - \frac{L}{2} \right )\leq \frac{4 D c_0 }{E_0} t,
\end{eqnarray}
again nearly linear, provided $R \gg L$ -- cf.\ Fig.\ \ref{fig:diffusive}.

\section{Experimental regimes}
The above considerations  indicate that the growth behavior of rod-like cells  depends strongly on whether their  nutrient capture rate is  reaction- or diffusion-limited.  Here, I estimate the nutrient concentration scale that separates these two regimes.   I make use of the fact that typical growing bacteria\footnote{Typical endogenous (non-growing) bacteria only consume energy at a rate of about  $1 \frac{\text{W}}{\text{kg}}$, two orders of magnitude smaller than the rate for typical growing cells \cite{makarieva2005energetics}.} consume energy at a rate of about $100 $ W/kg \cite{makarieva2005energetics}.  Assuming a cellular mass density of $1 \text{g}/ \text{cm}^3$ and a cell radius of $a \approx \frac{1}{2} \mu$m,  this corresponds to a sucrose influx per unit cellular length of\footnote{Sucrose has a molecular weight of 340 $\frac{\text{g}}{\text{mol}}$ and provides  $4 \frac{ \text{kcal}}{\text{g}}=5 \times 10^6 \frac{\text{J}}{\text{mol}}$.  In water, its diffusion coefficient is $5 \times 10^{-6}$ cm$^2$/s \cite{Berg-93}.}
\begin{eqnarray}\label{fluxin}
\frac{\mathcal{P}_{in}}{L} \sim \frac{3}{2} \times  10^{-20} \ \frac{\text{mol (sucrose)}}{\mu\text{m s}}.
\end{eqnarray}
Equating  this with the three-dimensional diffusion-limited influx per unit length from  (\ref{longlim}) gives  
\begin{eqnarray} \label{critc}
c_0 \sim \frac{ \log \left (\frac{L}{a} \right)}{2} \times 10^{-8} \text{ M (sucrose)}.
\end{eqnarray}
The expression (\ref{critc}) gives the  (weakly $L$-dependent) concentration of sucrose needed for the  diffusive influx into an isolated cell to just equal the rate at which growing bacteria consume nutrients in typical experiments:  Because experiments are typically carried out at high nutrient concentrations (cf.\  below), we can interpret (\ref{fluxin}) as a typical reaction-limited rate.  This implies that nutrient capture will be diffusion-limited below the concentration specified in  (\ref{critc}), and  reaction-limited above.

We can see that the interpretation above is self-consistent by considering the nutrient conditions typical of laboratory experiments.  Standard agar gel preparations might contain, for example, about 20 g/L of Bovril.  This corresponds to an energy content of about $10^5$ J/L, which is around six orders of magnitude larger than the energy concentration corresponding to (\ref{critc}): Growth in typical gels is thus indeed well into the reaction-limited nutrient capture regime\footnote{Notice from (\ref{twodinflux}) and (\ref{one-dres}) that $\mathcal{P}_{in}/L$ is almost inversely proportional to $L$ for diffusion-limited growth in one and two dimensions. This  strong $L$-dependence results in diffusion-limited growth at relatively high nutrient concentrations for these geometries.}.  This conclusion could also have been inferred from fact that  rod-like cells are almost always observed to grow at exponential rates in laboratory experiments  \cite{koch2001bacterial}, which we now know implies reaction-limited growth.
 
\section{Discussion}
Maintenance of a  steady rate of cellular growth requires the coordination of many different regulatory processes, some bio-chemical in nature \cite{klumpp2009growth, scott2010interdependence, Cro-pc2}, and others mechanical \cite{ami-13, hua-08,mukhopadhyay2009curvature,  fur-11, Ami-12, Nel-13}.  Here, I have not addressed these specific processes, but have instead attempted only to identify the running upper bounds on cell size that are set by the rate of nutrient capture.  Under reaction-limited growth, this approach returns an exponential upper bound on the running length of a rod-like cell, consistent with experimental observations.  However, the cell's shape, the nutrient's dynamics, and the environmental geometry each sensitively affect the maximum possible growth rate of a cell, and nutrient conservation generally leads to a sub-exponential growth bound.

   The above qualitative conclusions are entirely fundamental.    However, in order to derive simple, illustrative formulae, I have here made use of some simplifying approximations. One worth commenting on further is the nutrient steady-state approximation, equivalent to assuming that (\ref{diffeqn}) holds at all times.  As the cell grows faster and faster, this condition must eventually break down.  To estimate the cell length scale where this occurs, note that if the nutrient is to remain in steady-state, it must be able to diffuse over the relevant length scale, $L$, within one doubling period.  This is only  possible while
$L \partial_t L \lesssim O(D)$, a result that also follows from dimensional analysis. Now, a length doubling time of roughly $2000$ s  is exhibited in the data shown in Fig.\ \ref{fig:ellipse}.  Using a diffusion coefficient of $5 \times 10^{-6}$ cm$^2$/s, the value for sucrose in water \cite{Berg-93}, this returns a steady-state breakdown length of 
\begin{eqnarray}
L \sim \sqrt{2000 \log_2 e \times 5 \times 10^{-6}} \approx 0.1 \text{ cm}.
\end{eqnarray}
This is roughly $500$ times the typical cell length of $\approx 2$ $\mu$m, so the steady-state approximation should typically provide an accurate bound. Further,  the breakdown length scale is proportional to $\propto c_0^{-1/2}$, so  larger cell sizes can be considered by moving to smaller nutrient concentrations.

Connection to experiment can be made through comparison to the theoretical bounds on $\partial_t \log L  \equiv \frac{\gamma}{E_0}$, the rate of change of the logarithm of cell length.  It is important to point out that these bounds apply whenever a single nutrient limits cell growth:  This nutrient need not represent a source of energy, as I have assumed above; all that is needed is that the nutrient be the sole source of some essential material, inaccessible to cell through other means.  In an abundant or flowing nutrient ensemble,  (\ref{exp}) implies  $\frac{\gamma}{E_0} \leq \frac{\alpha_{in} 2 \pi a}{E_0}$, allowing for bounds on $\frac{\alpha_{in}}{E_0}$ to be obtained from measured growth curves.  If $\alpha_{in}$ can be estimated independently, this allows for $E_0$ to be bounded:  For example, using (\ref{fluxin}) and the doubling time $2000$ s exhibited by the data in Fig.\ \ref{fig:ellipse}, we obtain the following estimate for the energy needed to construct a micron length of  \textit{E.\ coli}:  $E_0 \lesssim 1.5 \times 10^{-10}$ J/$\mu$m.  Similarly, cellular metabolic rates can be determined through identification of conditions where $\mathcal{P}_{in} = \mathcal{P}_{m}$.  Whereas sensitive tuning of $c_0$ is needed to access this condition in a flow cell experiment,  an asymptotic, long-time bound $L^*$ on cell length always exists if growth is diffusion-limited.  In this case, in three, two, and one dimensions, setting $\mathcal{P}_{in} = \mathcal{P}_m$ returns $\alpha_M \sim \frac{2 D c_0}{a^2 \log \left ( \frac{L^*}{a} \right ) }$, $\frac{2 D c_0}{a^2 L^* \log \left (\frac{4R}{L^*} \right )}$, and $\frac{2 D c_0}{\pi a^2 R L^*} $, respectively. 
Measurement of $L^*$ should thus provide a relatively convenient and general method for estimating $\alpha_m$ values.

I turn now to some final, speculative comments relating to potential biological consequences of my results.  First,  recall that the influx of nutrient goes as the integral of $D \nabla c$ over the surface.  This gradient is the analog of the electric field in electrostatic systems, and the field near a conductor is highest at points of high surface curvature.  In the present context, this means that the influx of nutrient will be highest near the ends of the cell when diffusion-limited.  It is possible that end-growing bacteria have evolved to take advantage of this fact.  This possibility is also hinted at  by  observations showing that elongated cells often distribute membrane receptors accordingly, with higher densities near their tips \cite{rangamani2013decoding}.   Next I note that, from an energy consumption perspective, cell division provides no benefit in a homogeneous, flowing nutrient ensemble.  However, in a homogeneous, diffusion-limited ensemble, the power influx decreases steadily with $L$.  This suggests that maintenance of a nearly-optimal $\mathcal{P}_{in}$ --  which requires small cell size --  may represent a key driving force for division of rod-like cells. 
As a final point, I note that the nutrient-conservation analysis considered here for single cells is also applicable to groups of cells.  Collections of cells could in principle form any culture geometry desired, e.g., fractals, and this might lead to some interesting, exotic growth scaling laws.  It would be interesting to study whether or not collections of cells, initially disordered, ever  self-organize into configurations optimized for collective growth (e.g., rod-like aggregates that allow for exponential growth).  Many species of bacteria  (streptococci) and fungi form long chains, consistent with this possibility.  Considerations along these lines suggest that it would also be interesting to see if drugs could be designed to suppress the elongation of individual bacteria, or, alternatively, to break up chains of bacteria that cause infection.

\acknowledgements
I thank K.\ C.\ Huang and T.\ Ursell for many helpful discussions, two anonymous referees, A.\ Amir,  K.\ Mandadapu, D.\ B.\ McIntosh, and P.\ Rangamani for helpful comments and suggestions, the authors of \cite{ami-13} for kindly sharing the data shown in my Fig.\ \ref{fig:ellipse}, and D.\ Chandler for guidance and support.   This work was partially funded by a grant from the USA NSF: Grant No.\ CHE-1265664.

\section{Appendix: Derivation of diffusion-limited forms}
Here, I outline the solution to (\ref{diffeqn}) for a three-dimensional ellipsoidal cell.  The one- and two-dimensional cases can be treated in a similar manner.  For the ellipsoidal geometry, it is simplest to work in prolate spheroidal coordinates \cite{Arf-70}:  These are given by $(\sigma, \tau, \phi)$, where $\vert \textbf{r}_1\vert +\vert \textbf{r}_2 \vert \equiv 2 f \sigma$,  $\vert \textbf{r}_1\vert -\vert \textbf{r}_2 \vert \equiv 2 f \tau$, and $\phi$ is the azimuthal angle (the definitions of $f$ and the $\{\textbf{r}_i\}$ are given in Fig.\ \ref{fig:ellipse}).  Using these coordinates, (\ref{diffeqn}) separates and reads
\begin{eqnarray}\nonumber \label{3ddiff}
\frac{1}{f^2 (\sigma^2 -\tau^2)} \left  \{\partial_{\sigma} \left [(\sigma^2 -1) \partial_{\sigma}c \right]   + \partial_{\tau}\left [(1-\tau^2)\partial_{\tau}c \right] \right \} \\  + \frac{1}{f^2 (\sigma^2 -1)(1 -\tau^2)}\partial_{\phi}^2 c = 0.
\end{eqnarray}
The boundary condition at the cell's surface is $c\left (\sigma = \frac{L}{2f} \right) = 0$.  This can be satisfied only when the concentration is independent of $\tau$ and $\phi$ \cite{Arf-70}.  Seeking such a solution,  (\ref{3ddiff}) simplifies to $\partial_{\sigma} \left[(\sigma^2 -1) \partial_{\sigma}c\right]=0$.  Directly integrating this simpler equation gives
\begin{eqnarray}
c = c_0  - \frac{Q}{2f} \log \left  ( \frac{\sigma +1}{\sigma-1}\right),
\end{eqnarray}
which is equivalent to (\ref{exactc}).  The value of $Q$ given in (\ref{exactq}) is obtained by plugging into the inner boundary condition.

\bibliography{refs}

\end{document}